\documentclass[a4paper,fleqn,usenatbib]{mnras}

\usepackage{newtxtext,newtxmath}

\usepackage[T1]{fontenc}
\usepackage{ae,aecompl}

\usepackage{graphicx}	
\usepackage{amsmath}	
\usepackage{amssymb}	

\usepackage{enumerate}
\usepackage[british]{babel}

\title[]{An asteroseismic view of the radius valley:\\ stripped cores, not born rocky}

\author[V. Van Eylen et al.]{V. Van Eylen$^{1}$\thanks{E-mail: vaneylen@strw.leidenuniv.nl},
Camilla Agentoft$^{2}$,
M. S. Lundkvist$^{2,3}$,
H. Kjeldsen$^{2}$,\newauthor
J. E. Owen$^{4}$,
B. J. Fulton$^{5}$,
E. Petigura$^{5}$,
I. Snellen$^{1}$
\\
$^{1}$Leiden Observatory, Leiden University, postbus 9513, 2300RA Leiden, The Netherlands\\
$^{2}$Stellar Astrophysics Centre, Department of Physics and Astronomy, Aarhus University, Ny Munkegade 120, DK-8000 Aarhus C, Denmark\\
$^{3}$Zentrum f\"{u}r Astronomie der Universit\"{a}t Heidelberg, Landessternwarte, K\"{o}nigstuhl 12, 69117 Heidelberg, Germany\\
$^{4}$Astrophysics Group, Imperial College London, Blackett Laboratory, Prince Consort Road, London SW7 2AZ, UK\\
$^{5}$California Institute of Technology, Pasadena, California, USA\\
}

\date{Accepted XXX. Received YYY; in original form ZZZ}

\pubyear{2017}

\begin{document}
\label{firstpage}
\pagerange{\pageref{firstpage}--\pageref{lastpage}}
\maketitle

\begin{abstract}
Various theoretical models treating the effect of stellar irradiation on planetary envelopes predict the presence of a radius valley: i.e.\ a bimodal distribution of planet radii, with super-Earths and sub-Neptune planets separated by a valley at around $\approx 2~R_\oplus$. Such a valley was observed recently, owing to an improvement in the precision of stellar, and therefore planetary radii. Here we investigate the presence, location and shape of such a valley using a small sample with highly accurate stellar parameters determined from asteroseismology, which includes 117 planets with a median uncertainty on the radius of 3.3\%. 
~
We detect a clear bimodal distribution, with super-Earths ($\approx 1.5~R_\oplus$) and sub-Neptunes ($\approx 2.5~R_\oplus$) separated by a deficiency around $2~R_\oplus$. We furthermore characterize the slope of the valley as a power law $R \propto P^\gamma$ with $\gamma = {-0.09^{+0.02}_{-0.04}}$. A negative slope is consistent with models of photo-evaporation, but not with the late formation of rocky planets in a gas-poor environment, which would lead to a slope of opposite sign. The exact location of the gap further points to planet cores consisting of a significant fraction of rocky material.
\end{abstract}
  



\section{Introduction}

Various theoretical models predict that planets at short orbital periods are strongly influenced by the radiation of their host stars. For example, at the shortest orbital period a ``photoevaporation desert'', i.e.\ an absence of sub-Neptune-size planets $(1.8-4.0\,R_\oplus)$ and an increase in rocky planets $(R{<}1.8\,R_\oplus)$ has been predicted \citep{lopez2013} and observed with increasing clarity as the precision of stellar parameters increased \citep{borucki2011,lundkvist2016}. 

Furthermore, formation models predict that atmospheric erosion of short-period planets results in the presence of a ``photoevaporation valley'', i.e.\ a gap in the radius distribution of planets around $1.75-2~R_\oplus$ \citep{owen2013,jin2014,lopez2014,chen2016,lopez2016,owen2017}. This valley defines the boundary between planets with a mass large enough to hold on to their gas envelope, and planets which have been stripped of their atmospheres and consist of the remnant core. The specific shape and slope of the valley depends on the details of planet formation, the composititon of the formed planets and the physics of evaporation \citep[e.g.][]{lopez2016,owen2017}.

Observing this valley is not straightforward and is complicated by the relatively high uncertainty in observed planet radii, a result of uncertain stellar radii \citep{owen2013}. Recently, \cite{fulton2017} provided clear evidence of the valley by using a spectroscopic sample from the California-\textit{Kepler} Survey (CKS), with better-constrained stellar parameters \citep{petigura2017,johnson2017}. Despite the clear detection of the bimodal radius distribution and a radius gap, \cite{fulton2017} did not attempt to constrain the slope of this gap as a function of orbital period.

Here, we investigate the radius gap using a sample with homogeneously determined stellar parameters from asteroseismology \citep{silvaaguirre2015,lundkvist2016}. This sample is smaller than the CKS sample, but has better constrained stellar parameters, which translate into more accurate planet parameters.  

In Section~\ref{sec:methods} we describe our sample and parameter determination. In Section~\ref{sec:radiusgap} we show the modeling of the radius valley. Finally, in Section~\ref{sec:models} we compare our findings with theoretical predictions, and we draw conclusions in Section~\ref{sec:conclusions}.

\section{Methods}
\label{sec:methods}	

In this work, we combine accurate stellar parameters from asteroseismology \citep{silvaaguirre2015,lundkvist2016} with carefully modeled planet transits, to investigate the location, size, and shape of the so-called `radius gap'. We first detail how we determine planet parameters, and then describe the properties of our sample. 

\subsection{Parameter Determination}
\label{sec:sample}

To determine accurate planet parameters from transit surveys, accurate stellar parameters are required, because the transit depth only constrains $R_\mathrm{p}/R_\star$, where $R_\mathrm{p}$ and $R_\star$ are the planetary and stellar radius, respectively. We therefore start from a sample of exoplanet host stars with parameters homogeneously measured from asteroseismology, which can provide highly precise masses and radii for a sample of bright stars. For systems with multiple transiting planets, we use the planet modeling by \cite{vaneylen2015}, which uses stellar parameters taken from the asteroseismic modeling by \cite{huber2013} and \cite{silvaaguirre2015}. For systems with a single transiting planet, planet modeling was similarly done by \cite{vaneylen2017}, which uses the slightly more complete asteroseismic catalogue by \cite{lundkvist2016}. Those asteroseismic catalogues are fully consistent \citep{lundkvist2016}. 

We summarize the planet modeling approach here. We start from the Presearch Data Conditioning (PDC) data \citep{smith2012}. Using an iterative approach, the times of individual transits are determined using the transit model parameters. The individual transit times are then used to determine the best orbital period, and determine if any transit timing variations (TTVs) are present. The systems for which a sinusoidal TTV model is included are detailed in \cite{vaneylen2015} and \cite{vaneylen2017}. Planet transits are modeled with analytical transit equations \citep{mandel2002}. Our fitting procedure uses a Markov Chain Monte Carlo (MCMC) aproach using the $emcee$ code \citep{foremanmackey2013}, a Python implementation of the Affine-Invariant Ensemble Sampler \citep{goodman2010}. Eight planet parameters are sampled, namely the ratio of planet to star radius ($R_\textrm{p}/R_\star$), the impact parameter ($b$), two combinations of eccentricity and angle of periastron $e$ and $\omega$ ($\sqrt{e}\cos \omega$ and $\sqrt{e} \sin \omega$), the time of mid-transit ($T_0$), an offset in flux ($F$), and two stellar limb darkening parameters following a quadratic limb darkening law ($u_1$ and $u_2$). A flat prior is used for all parameters except limb darkening, for which a Gaussian prior was used, with the mean value predicted from a Kurucz atmosphere table \citep{claret2011} and a standard deviation of 0.1. The stars are cross-checked for contamination from nearby stars from high-resolution imaging \citep[e.g.][]{furlan2017}. We refer to \cite{vaneylen2015} and \cite{vaneylen2017} for a more detailed description of the transit analysis method. The stellar mass and radius, and the planet radius and orbital period are listed in Table~\ref{tab:parameters} for all systems in our sample.

\subsection{Sample Properties}
\label{sec:analysis}

As a starting point, we use the sample of planet host stars with homogeneously-determined asteroseismic parameters \citep{huber2013,silvaaguirre2015,lundkvist2016}. As detailed in \cite{vaneylen2015} and \cite{vaneylen2017}, a few systems were removed from the initial sample, e.g.\ because they have subsequently been identified as false positives or likely false positives, or because they have not been observed in \textit{Kepler}'s one minute short cadence sampling, which decreases the precision of the derived stellar and planetary parameters. Most of the planets have been confirmed or validated, while 17 are unconfirmed planet candidates that are likely to be bona fide planets \citep{morton2016,vaneylen2017}. The final sample contains 75 stars and 117 planets, which are listed in Table~\ref{tab:parameters}. 

\begin{figure*}
	\includegraphics[width=2\columnwidth]{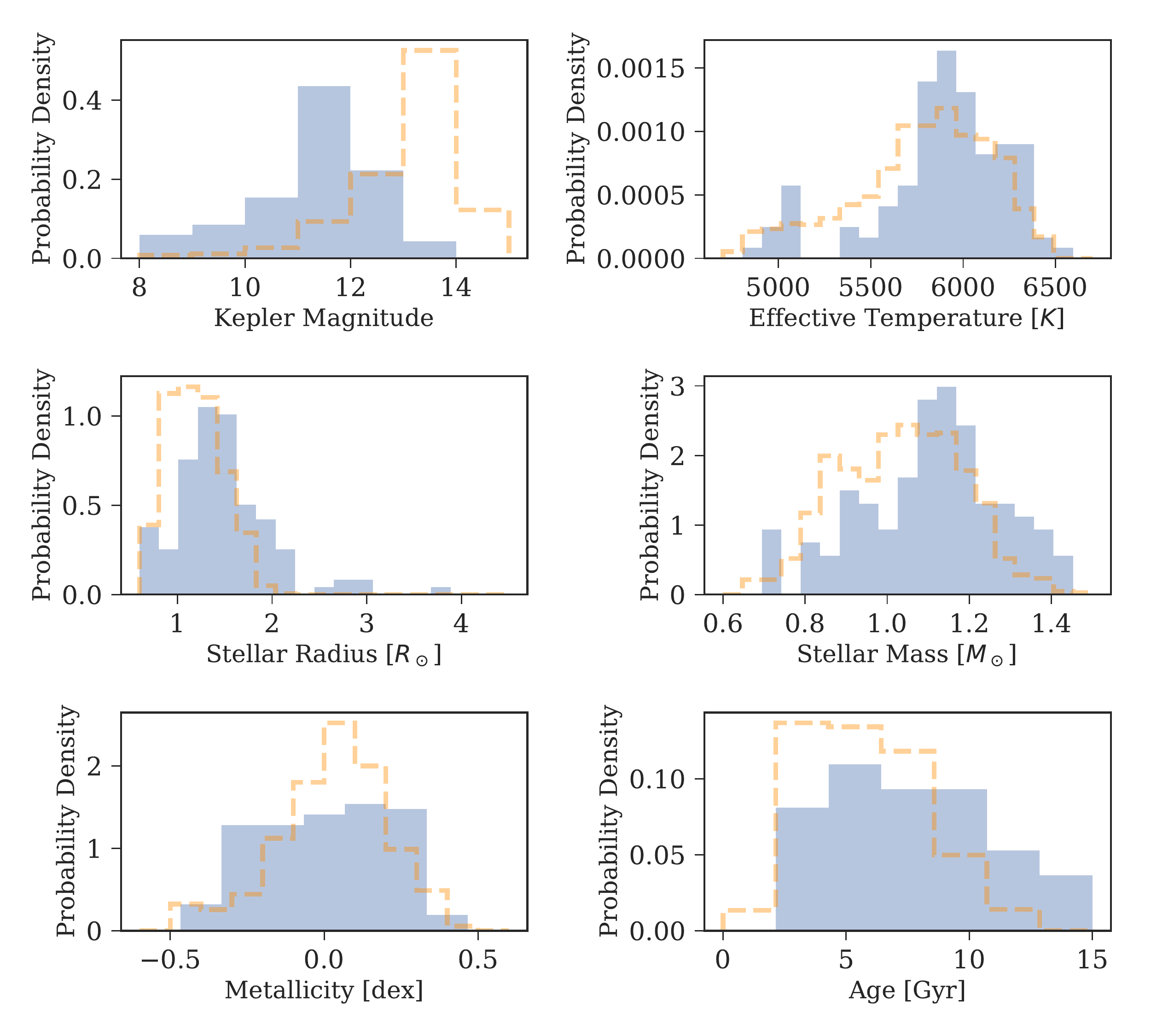}
	\caption{Histograms showing the basic properties of our sample in blue, and the properties of the sample by \protect\cite{fulton2017} in orange. The asteroseismic sample contains brighter stars. The stellar parameter range is similar, but on average this sample contains brighter, larger and older stars. Stars with multiple planets are counted multiple times, but the shape of the histogram is not fundamentally changed if each star is only counted once.}
	\label{fig:starhistogram}
\end{figure*}

A histogram of the \textit{Kepler} magnitude, stellar effective temperature, stellar radius, stellar mass, metallicity, and age is shown in Figure~\ref{fig:starhistogram}. The parameters for the multi-planet systems are taken from \cite{vaneylen2015}, with stellar parameters originally from \cite{huber2013} and \cite{silvaaguirre2015}, and those for the single planet systems are from \cite{vaneylen2017} with stellar parameters from \cite{lundkvist2016}. The ages, when available, were taken from the BASTA pipeline \citep{silvaaguirre2015}.

The stellar properties of this sample are driven by the requirement of measurable p-mode oscillations, resulting in a sample containing primarily bright stars with the average star being slightly more massive and slightly more evolved than the Sun. The mean \textit{Kepler} magnitude is 11.3. In Figure~\ref{fig:starhistogram}, we compare the properties of our sample with that of \cite{fulton2017} using the stellar parameters from \cite{petigura2017}. The stellar properties of our sample are broadly similar to those investigated by \cite{fulton2017}, which contains main sequence stars with a temperature range of 4700-6500~K. Our sample spans only the bright end of the \cite{fulton2017} stars and is significantly smaller -- 117 planets, compared to 900 in the adopted \cite{fulton2017} sample.
Compared to the \cite{fulton2017} sample, the average star in this sample is larger and older. 
This is a consequence of the selection for solar-like oscillations, which are easier to detect in more evolved stars due to their larger oscillation amplitudes, but there are no obvious biases which would affect the distribution of planet parameters.

The parameters in this sample are determined to significantly greater precision, e.g.\ the median fractional uncertainty on the stellar radius is 2.2\%, or $0.03~R_\odot$, which can be compared to an 11\% uncertainty in the CKS sample \citep{fulton2017} and a 25\% uncertainty in the more general \textit{Kepler} catalogue \citep{huber2014}. This, in turn, leads to a median fractional uncertainty on the planet radius of 3.3\% (or 0.068 $R_\oplus$), compared to 12\% in the CKS analysis \citep{fulton2017}.

\section{Radius-Period Gap}
\label{sec:radiusgap}

\begin{figure}
	\includegraphics[width=\columnwidth]{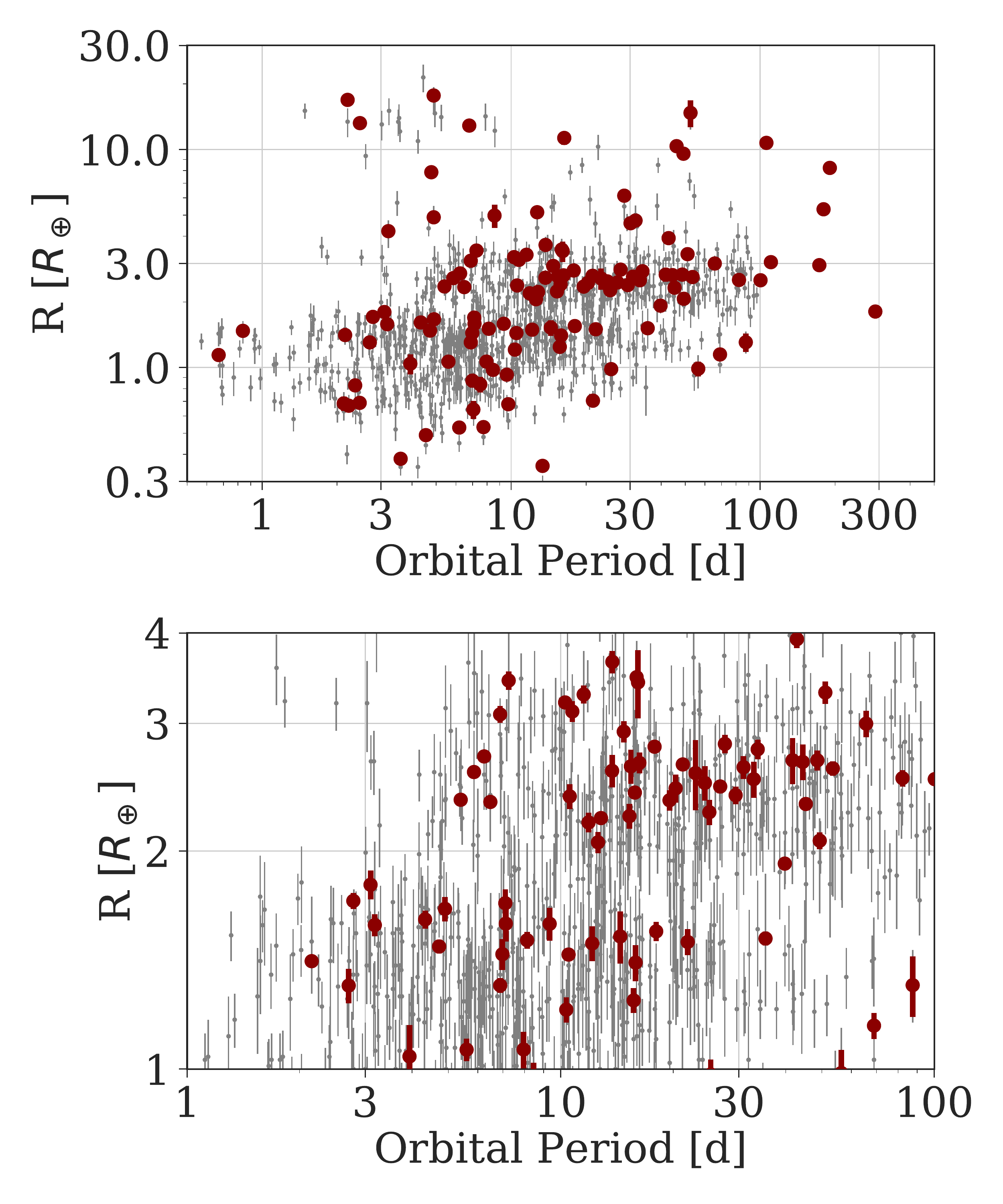}
	\caption{The planet radius as a function of orbital period. In grey, data points and uncertainties by \protect\cite{fulton2017} are shown, while the sample described here is shown in red. In many cases, the uncertainties are smaller than the symbol size. The bottom plot highlights the part of the sample where the radius gap occurs, at $R \approx 2~R_\oplus$.}
	\label{fig:periodradius}
\end{figure}

\begin{figure}
	\includegraphics[width=\columnwidth]{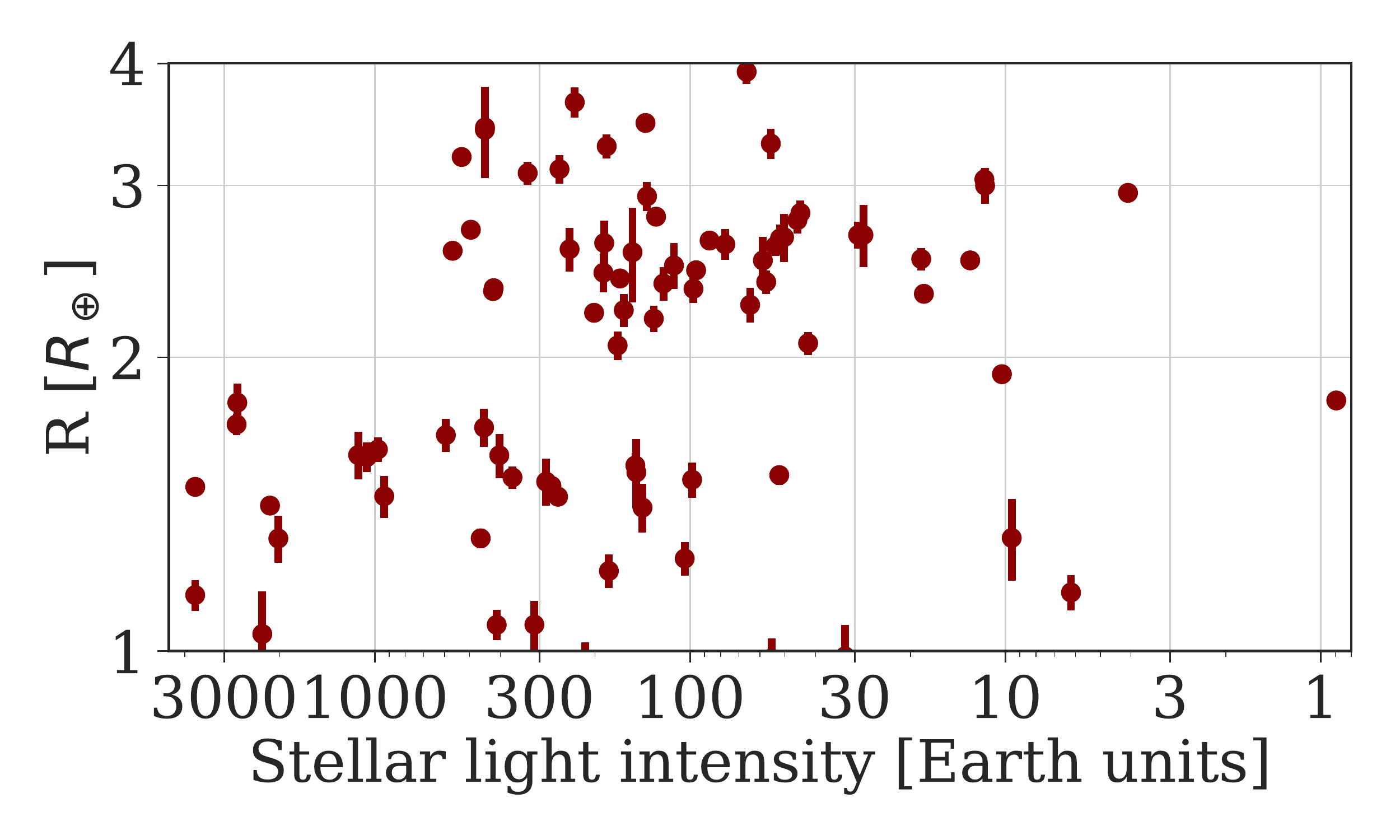}
	\caption{Similar to Figure~\ref{fig:periodradius}, but with the planet radius as a function of incident flux rather than orbital period. In many cases, the uncertainties are smaller than the symbol size. The x-axis has been inverted, so that high incident flux (short orbital periods) are on the left. As before, only planets smaller than 4~$R_\oplus$ are shown.}
	\label{fig:fluxradius}
\end{figure}

The planets in our sample are plotted in a period-radius plane in Figure~\ref{fig:periodradius}, and compared to the sample by \cite{fulton2017} which is larger but has higher uncertainties. We also plot the sample as a function of incident flux in Figure~\ref{fig:fluxradius}.

\begin{figure}
	\includegraphics[width=\columnwidth]{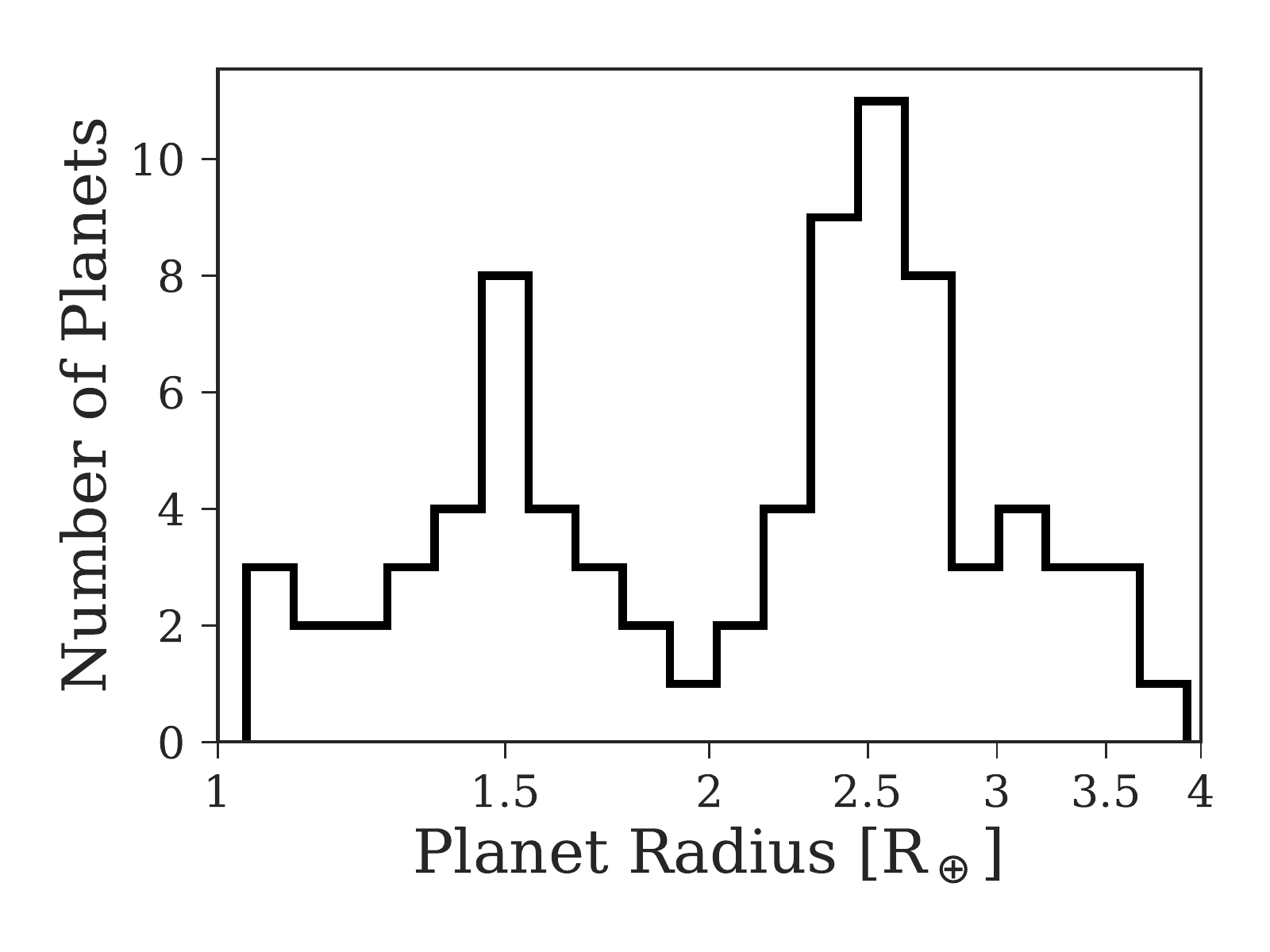}
	\caption{A histogram of the number of planets in the sample as a function of planet radius, with $1~R_\oplus \leq R \leq 4~R_\oplus$, using 20 logarithmic radius bins. Two peaks can be observed at $\approx 1.5~R_\oplus$ and $\approx 2.5~R_\oplus$, with a low density of planets in between.}
	\label{fig:radiushistogram}
\end{figure}

We now limit our sample to planets smaller than $4~R_\oplus$. Even by eye, an absence of planets around $R \approx 2~R_\oplus$ can be seen. 
In Figure~\ref{fig:radiushistogram}, we show a histogram of the planet radius, which similarly shows a bimodal distribution with peaks roughly at $\approx 1.5~R_\oplus$ and $\approx 2.5~R_\oplus$, and a dip in between these peaks. 

Figure~\ref{fig:radiushistogram} has not been corrected for transit probability, which is slightly lower for the planets above the gap, which occur at longer average periods, and has furthermore not been corrected for detection probability, which is lower at the smallest planets which are more likely to be missed. These corrections would be important to calculate absolute planet occurrence, but the sparseness of our sample makes it poorly suited for this purpose. However, any such correction would not affect the bimodal shape of the histogram.

Furthermore, we investigate whether planets at longer orbital period (i.e. $P>25$~days) could be missed, due to their lower signal-to-noise ratio (SNR) than shorter period counterparts. All planets in our sample are detected at a high SNR. We can estimate the SNR through the fractional uncertainty on $R_\mathrm{p}/R_\star$, which is a measure for how well we can measure the transit depth. We find a mean SNR of 20 for planets with $R<2~R_\oplus$ with orbital periods between 25 and 100 days, indicating that even in this part of parameter space, we can detect planets at high significance.

We now constrain the shape and slope of the gap as a function of radius and orbital period. We first attempt to directly fit the absence of data points itself, using a linear model $\log R_\mathrm{mod} = m \log P_\mathrm{mod} + a$, where $R_\mathrm{mod}$ and $P_\mathrm{mod}$ are the modeled radius and period, and $m$ and $a$ are the slope and offset we set out to determine. To fit an absence of data points (the `gap'), we invert the likelihood function, i.e. 

\begin{equation}
 \log L = -0.5 \sum_i \frac{(\log R_\mathrm{obs} - \log R_\mathrm{mod})^{-2}}{\sigma_{\log R}^2} - \log{\frac{1}{\sigma_R^2}},
\end{equation}
where $R_\mathrm{obs}$ and $R_\mathrm{mod}$ are the observed and modeled planet radii, and $\sigma_R$ is the uncertainty on the observed radius. Here, the power $-2$ ensures that the fit maximizes the distance to observations, fitting an absence of data, rather than the usual factor $2$, when attempting to make a best fit through the observed data points. We then optimize the likelihood with an MCMC algorithm \citep[\protect\textit{emcee,}][]{foremanmackey2013}, using uninformative flat priors on the slope $m$ and offset $a$, while limiting their range to $-0.5 \leq m \leq 0.5$ and $\log 1 \leq a \leq \log 4$, to ensure that the fit remains within our range of observations. We fit all data with $R \leq 4~R_\oplus$, and $1 \leq P \leq 100$~days. We sample with 10 walkers, taking 4000 steps each, after a burn-in phase of 2000 steps.

We find median values $m = -0.08$ and $a = 0.34$. 
To ensure the slope of the gap is not affected by planets which could be missed at longer orbital periods, we further calculate the SNR of planets below the gap at $P>25$~days, and find a SNR of 17 for these planets, indicating transiting planets below the gap at relatively long period can be discovered at high significance in our sample of bright stars.
This suggests the slope is not driven by planet detectability at longer periods. However, we also model the slope after limiting our sample to $P \leq 25~$days. Here, we find $m = -0.10$ and $a = 0.35$, showing that our slope measurement is not the result of (a lack of) planets at longer orbital periods. Within the limitations of our sample, the measurement of the slope is largely independent of the precise period cut. A downside of this approach is that this likelihood function leads to unrealistically small uncertainties which depend heavily on the uncertainty of the observed radii. However, the true uncertainty of the slope of the radius valley is a result of the sparseness of the sampling, rather than the precision with which individual radii are measured. 

\begin{figure}
	\includegraphics[width=\columnwidth]{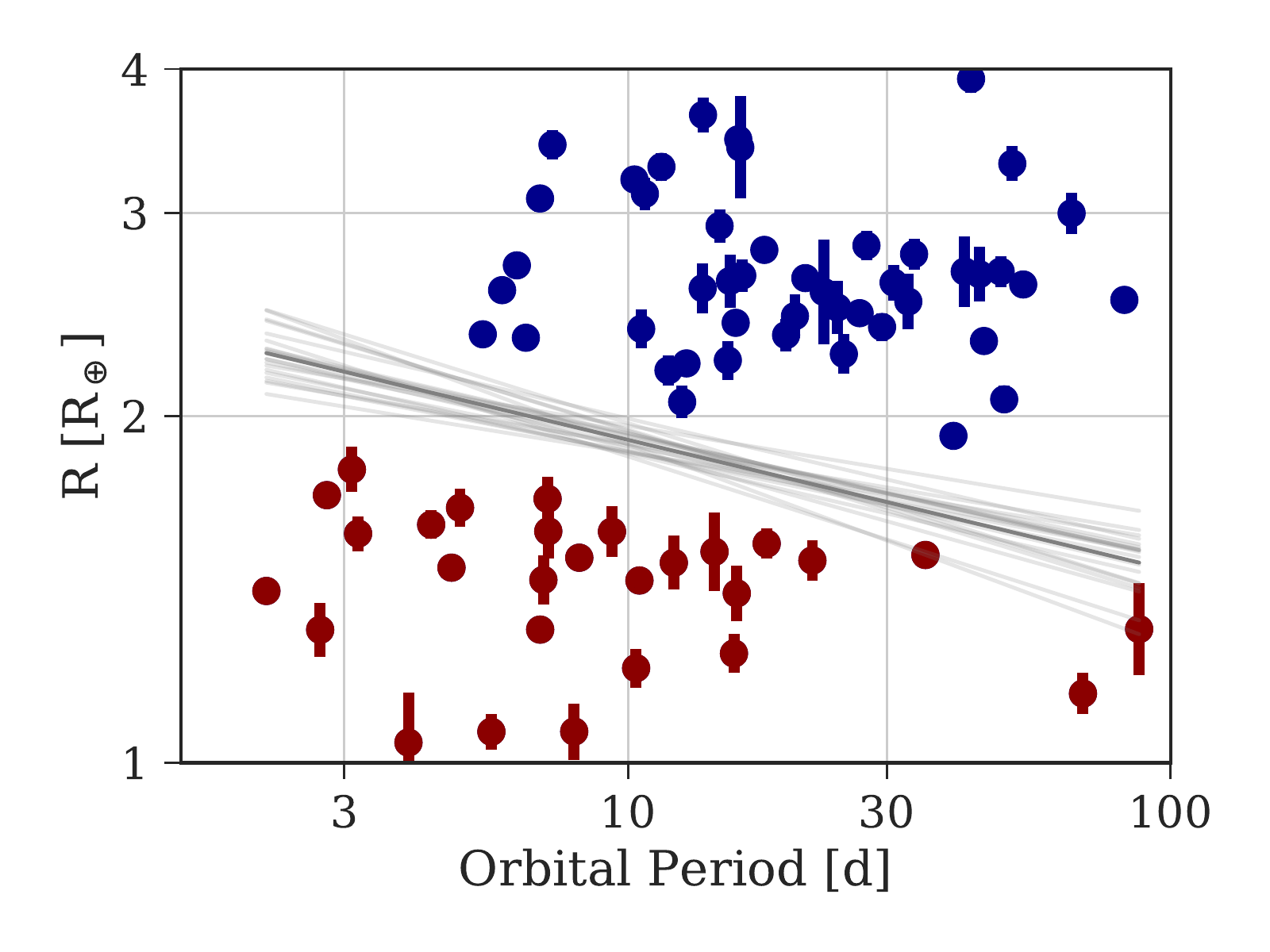}
	\caption{The grey lines show the best fits to the bootstrapped samples, 20 fits out of 1000 bootstrapped samples are shown, with the thicker line showing their average. We find a slope $m = 0.10 \pm 0.03$ and offset $a = 0.38 \pm 0.03$. We use these fits to separate our sample into planets below the gap (red) and planets above (blue).}
	\label{fig:gapfit}
\end{figure}

To calculate the uncertainty due to our sampling, we make bootstrap versions of our sample, by generating new samples with the same size from our observed sample, allowing replacement. In these new, bootstrapped samples, some planets will be counted multiple times, while others may not be counted at all. In this way, we generate 1000 new samples, and apply the MCMC algorithm to each of these, as described above. We then take the 50\% quantile for all samples of $m$ and $a$, and use the 16\% and 84\% quantiles for the uncertainties. We find $m = -0.10 \pm 0.03$ and $a = 0.38 \pm 0.03$, which as expected results in similar values, but with higher, more realistic uncertainties. In Figure~\ref{fig:gapfit}, we show 20 randomly drawn linear fits. 
We again ensure our result does not depend on orbital period and rerun our model after limiting the sample to $P < 25$~days. Here, we find $m = -0.13^{+0.04}_{-0.05}$ and $a = 0.41 \pm 0.05 $, which is a slightly steeper slope, albeit consistent at $1\sigma$ with the values above.

\begin{figure}
	\includegraphics[width=\columnwidth]{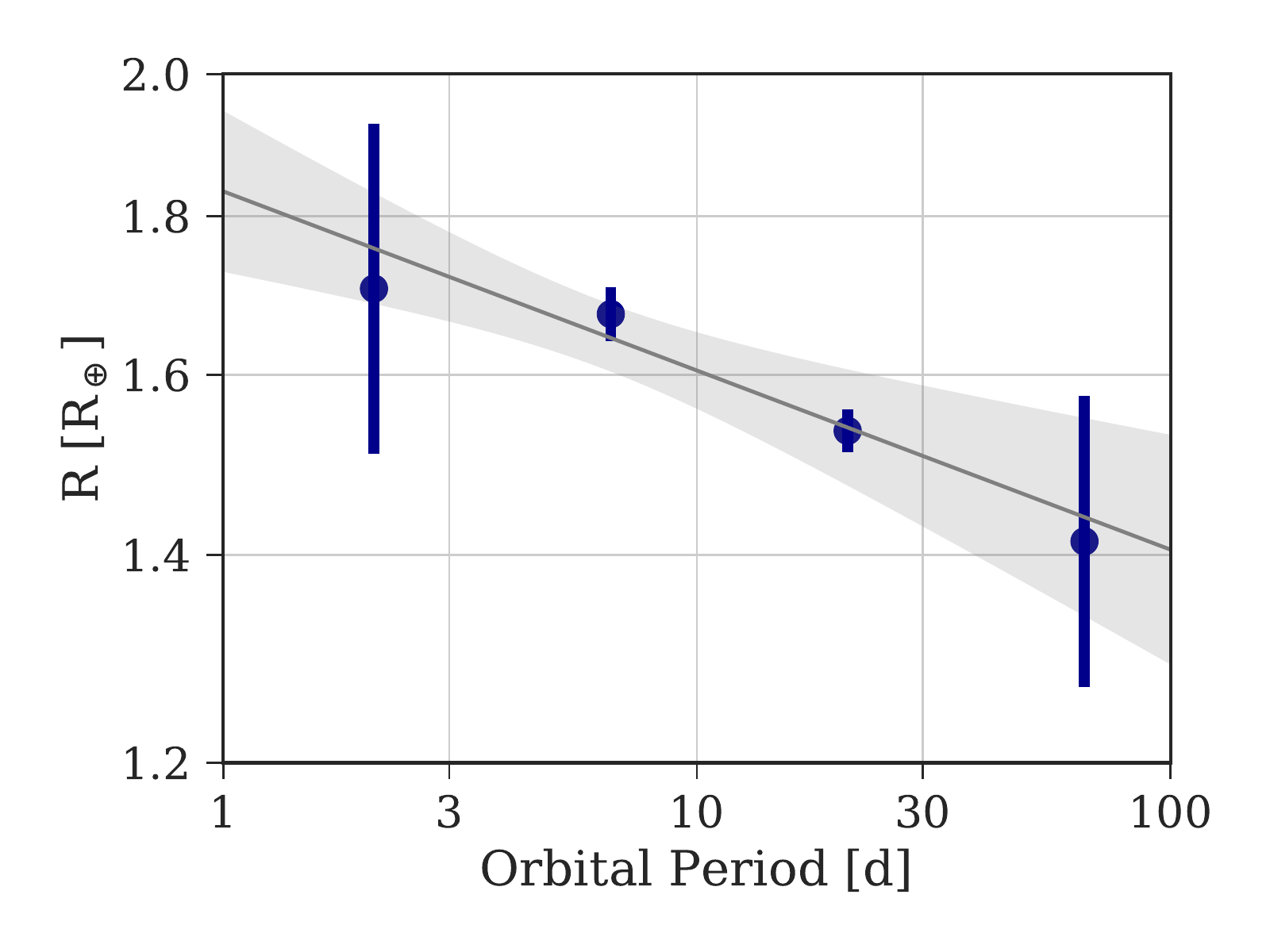}
	\caption{The data points show the maximum radius of planets below the gap as a function of orbital period, with the uncertainty derived from 1000 bootstrap iterations on the initial sample. The grey shows the best fit, together with a 68\% confidence interval, again derived from the bootstrap iterations. We find that the slope of the gap is $m = -0.05^{+0.01}_{-0.03}$ and $b = 0.26 \pm 0.02$.}
	\label{fig:maxfit}
\end{figure}

We can now use these fits to the gap to separate our sample into planets below and above the gap. Subsequently, we can look at the planets below the gap to directly estimate the slope of the gap, by looking at the maximum radius of these planets as a function of orbital period. We create four logarithmic bins as a function of period, and calculate the maximum radius in each bin, repeating the procedure by resampling our data, again allowing repetition of indivdual observations. We then apply a linear regression to each of the bootstrapped samples, and again calculate 16\%, 50\% and 84\% quantiles. We find that the slope of the maximum of the lower part of the radius valley, i.e.\ the super-Earths, is $m = -0.05^{+0.01}_{-0.03}$ and $b = 0.26 \pm 0.02$. The result is shown in Figure~\ref{fig:maxfit}.

\begin{figure*}
	\includegraphics[width=2\columnwidth]{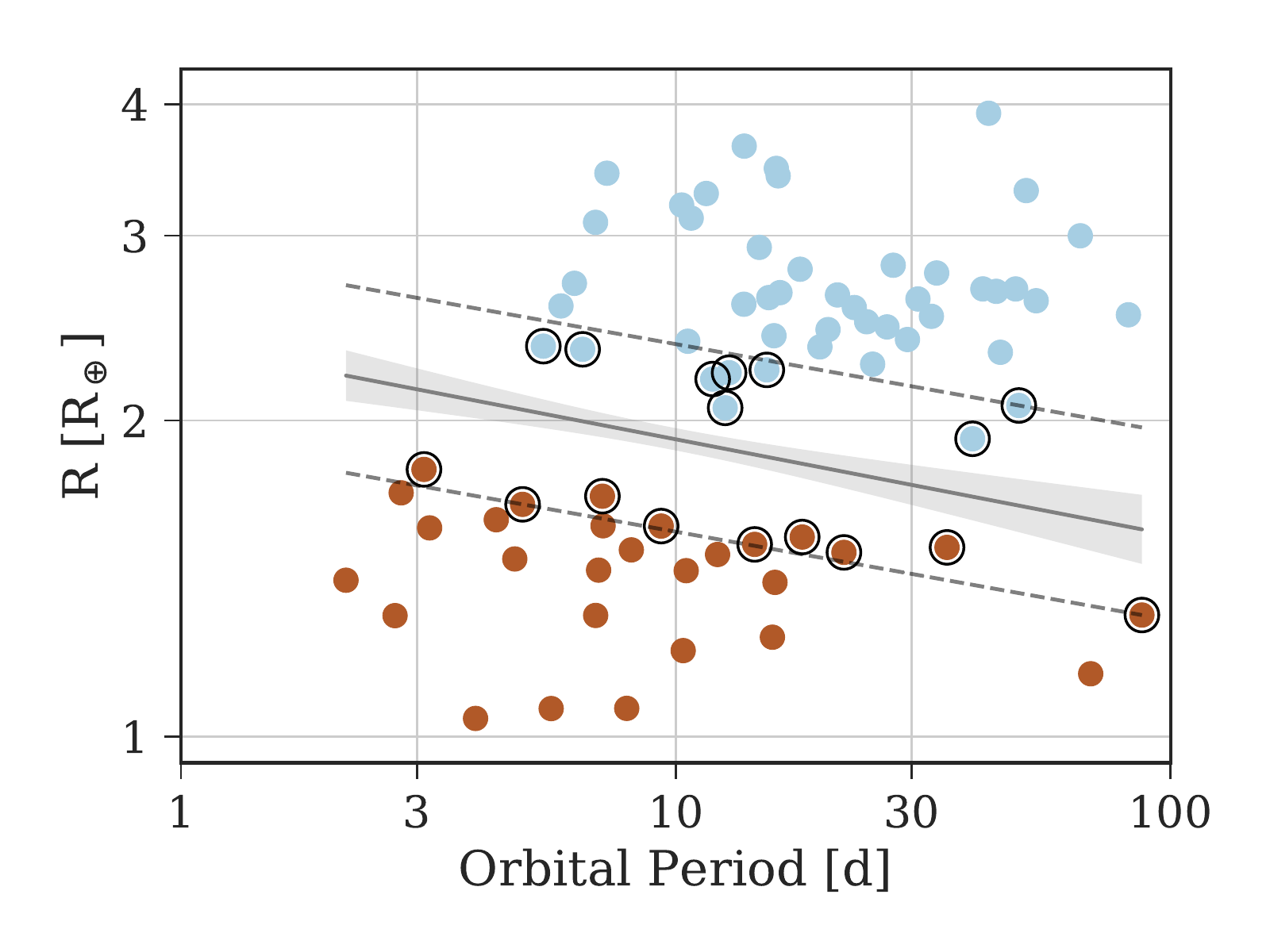}
	\caption{The slope of the radius valley as determined by support vector machines. The grey line represents the hyperplane of maximum separation, together with a 68\% confidence interval derived from bootstrapping the original sample. The super-Earths below the radius valley are shown in red, while the sub-Neptunes above the valley are plotted in blue. The encircled data points are the support vectors, which determine the slope of the radius valley. The parallel dotted lines go through the support vectors, and are determined by offsets $a_\mathrm{low} = 0.29^{+0.04}_{-0.03}$ and $a_\mathrm{upp} = 0.44^{0.04}_{-0.03}$ respectively.}
	\label{fig:svc}
\end{figure*}

The downside of this approach is that it uses only a few observations (i.e.\ none of the sub-Neptunes were included) and is potentially sensitive to binning. A more robust approach makes use of support vector machines to determine the hyperplane of maximum separation between the planets above and below the valley. This line of separation maximizes the distance to points of the different classes of data (in this case, the super-Earths below the valley, and the sub-Neptunes above). Here, we use the Python implementation of support vector classification, \textit{SVC}, in the scikit machine learning package \textit{sklearn}. To determine the hyperplane a penalty parameter $C$ has to be chosen. This parameter determines the trade-off between maximizing the margin of separation and the tolerance for misclassification of observations, with high values of $C$ allowing the lowest amount of misclassification. 

This suggests that in this case, we want to use a high value of $C$, because the data points in our sample are well-separated into super-Earths and sub-Neptunes, and we only want to use the data points close to the gap to determine its shape. Indeed, if we pick a low value of $C$ (e.g.\ $C=1$), almost all data points are used to separate the sample, and we find that this no longer fits the radius valley (a high degree of misclassification) and leads to a steep (negative) slope which no longer visually matches the observed valley. By contrast, picking a very high value for $C$ implies the hyperplane is determined by only very few support vectors (i.e.\ data points nearest to the valley). For example, for $C=100$, the hyperplane is determined by only four support vectors, i.e.\ two super-Earths and two sub-Neptunes, leading to $m = -0.08^{+0.02}_{-0.01}$, where the uncertainties were calculated using 1000 bootstrapped samples as before. We finally calculate the hyperplane of maximum separation using a compromise between these extremes, i.e.\ $C=10$. As can be seen in Figure~\ref{fig:svc}, using this value, the slope of the valley is determined by about 15 support vectors, i.e.\ 15 planets closest to it. This provides results consistent with the lower $C$ value above, but with more support vectors this leads to a larger uncertainty. Again following our bootstrapping approach, we find $m = -0.09^{+0.02}_{-0.04}$ and $a = 0.37^{+0.04}_{-0.02}$.  

In summary, in this section we have used different methods to determine the slope of the observed radius valley. All these methods find consistent and distinctly negative slopes. Because support vector machines provided the most standardized way of separating samples, we use these as our preferred parameters, although some readers may prefer to use one of the other methods, or calculate their own slope based on the parameters listed in Table~\ref{tab:parameters}.

\section{Discussion}
\label{sec:models}

We observe a bimodal distribution of planet radius, broadly peaking at $\approx~1.5~R_\oplus$ and $\approx~2.5~R_\oplus$, with a valley at around $1.7-2~R_\oplus$ in between. The radius valley has also been observed recently by \cite{fulton2017}. The feature we observe here is broadly similar, although the valley is more pronounced in our sample, presumably because the stellar and planetary radii are determined more accurately in this work. The \cite{fulton2017} sample is significantly larger, enabling a determination of occurrence rates of planets for different radii and periods, which is beyond the scope of this work. By contrast, owing to a highly precise asteroseismic sample of stellar parameters, we were able to measure the slope of the radius valley as a function of orbital period for the first time, and find $m = -0.09^{+0.02}_{-0.04}$. 

A large body of theoretical work predicts and interprets the existence of a planet occurrence valley as a function of planet radius and orbital period or incident flux. Even when planets form with a continuous distribution of initial radii, photoevaporation can produce a deficit of planets around $2~R_\oplus$ \citep{owen2013}. In such a model, planets can either maintain hydrogen envelopes, or not, depending on their XUV exposure, creating a bimodal distribution in planet sizes. Similarly, \cite{lopez2013} predict an occurrence valley with a width of roughly $0.5~R_\oplus$, occurring at larger radii for closer-in planets. 

The physical reason for a deficit or gap is that planets around this radius would have a very small envelope, which is stripped off easily, even at low mass-loss rates. The mass-loss timescale peaks when the envelope approximately doubles the radius of the planet. Planets with a smaller envelope are unstable to complete evaporation, because the mass-loss timescale decreases during evaporation. On the other hands, planets with a larger envelope see their mass-loss timescale increase as mass is removed, which stabilises when they are double the core radius. As a result, planets that resisted full photo-evaporation end up with substantial envelopes, which contribute significantly to the planet radius, and make up $\approx~1-10\%$ of their mass \citep{lopez2014}. Meanwhile, other planets end up with virtually no envelopes at all and remain as the stripped cores.

An alternative physical process to strip the atmosphere of some planets comes from the luminosity of the cooling rocky core itself \citep{ginzburg2018}, and would similarly produce a radius valley. Another mechanism that may explain the large diversity in mean density of short-period planets, is late giant impacts which lead to atmospheric erosion \citep[e.g.][]{liu2015,inamdar2016}. However, while this mechanism would influence the mass distribution of these planets, it is unclear how it could lead to a clear period valley. 

\cite{lopez2016} investigate the possibility that the short period super-Earths are a separate population of rocky planets which never had significant envelopes, rather than stripped cores of planets that lost their envelopes. This could occur if these planets formed after the proto-stellar disks had already evaporated, in a similar way as to how the Earth has likely formed. Understanding whether the short-period super-Earths are the result of photo-evaporation or are primordial rocky planets is therefore important to constrain the frequency of planets like Earth in the habitable zone \citep{lopez2016}. 

In the case of this late, gas-poor formation, the transition radius would be a function of the available solid material that a planet core can accrete due to collisions. This would result in a transit radius dependence on orbital period between $P^{0.07}$ and $P^{0.10}$, i.e.\ the radius valley increases with orbital period \citep{lopez2016}. This is in clear contrast with the photo-evaporation scenario. In that case, planets with the largest core mass are the most resistant to photo-evaporation, so that at short orbital periods, the transition radius is larger, and may scale with orbital period as $P^{-0.15}$ \citep{lopez2016}. Similarly, \cite{owen2017} find that the radius of the bottom of the valley depends on orbital period as $P^{-0.25}$ to $P^{-0.16}$, depending on the photo-evaporation model, and where the location of the valley depends on the properties of the remnant cores. Numerical models empirically give shallower slopes than analytic models for the same evaporation models, e.g.\ a slope of $P^{-0.12}$ is found from numerical models, for an analytical slope of $P^{-0.16}$ \citep{owen2013}.

\begin{figure}
\includegraphics[width=\columnwidth]{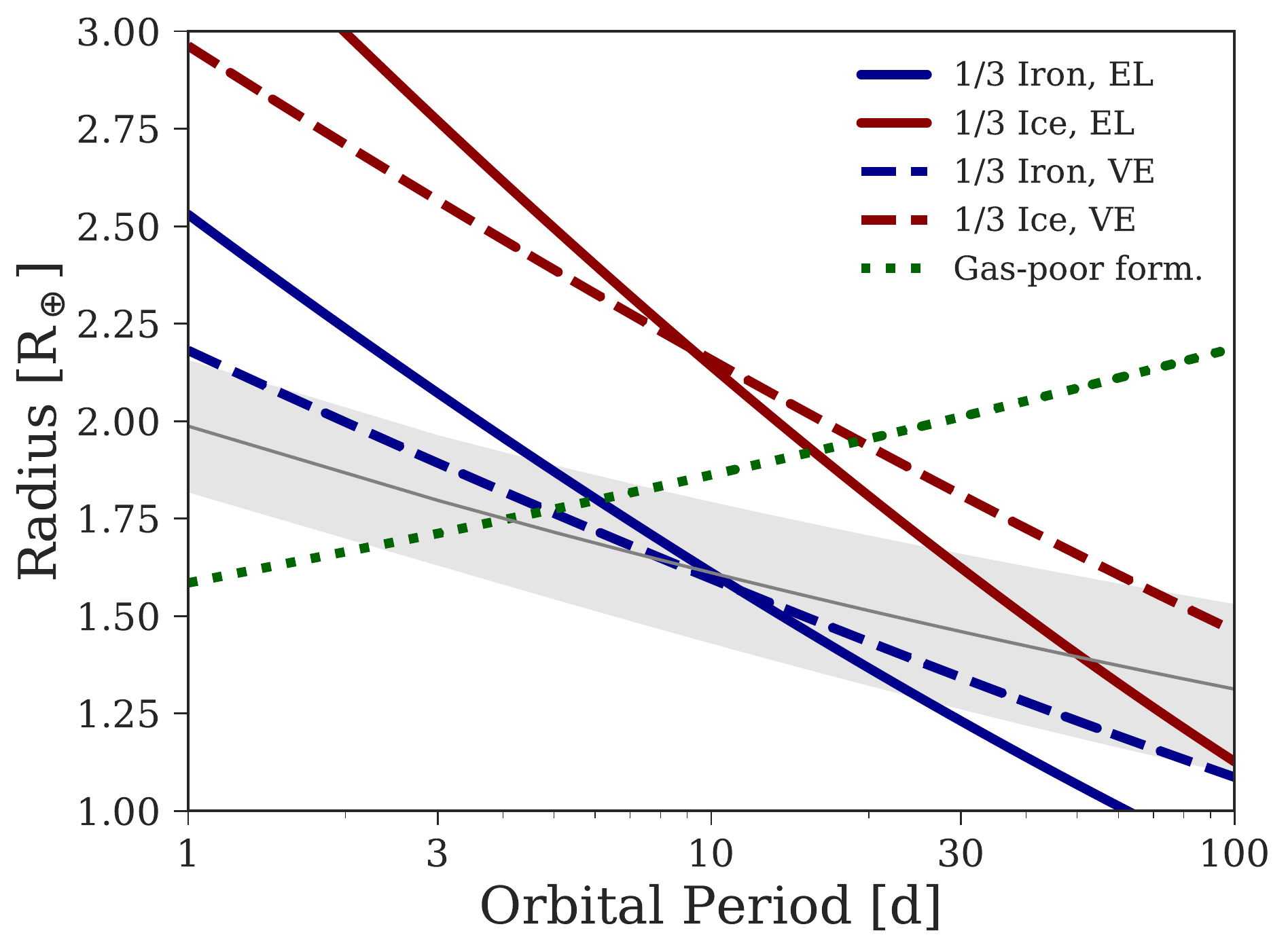}
	\caption{We compare the observed slope of the radius gap to theoretical models with different planet core compositions from \protect\cite{owen2017}, showing the position of the bottom of the evaporation valley, which is the largest super-Earth at a given orbital period. In grey, we show the best value and $1\sigma$ confidence interval from the support vector machine determination of the period valley, using the lower parallel line, shown in Figure~\protect\ref{fig:svc}. Different models for the bottom of the evaporation gap are shown, with solid lines showing a constant efficiency energy-limited (EL) models while dashed
lines show evaporation models with variable efficiency \protect\cite[VE, see e.g.][]{owen2012}. The blue lines show Earth-like composition cores (1/3 Iron). The red lines show planets which consist of 1/3 ice and 2/3 silicates. We find that our observations provide the best match with Earth-like cores and a variable efficiency. We refer the reader to \protect\cite{owen2017} for details about the models. In a dotted green line, we show the predicted slope of the radius valley in case of a gas-poor formation model, from \protect\cite{lopez2016}.
}
	\label{fig:theory}
\end{figure}

The negative slope we observe here is consistent with physical models of photo-evaporation, but not with late formation, in a gas-poor environment after the disc has dissipated, which would predict a slope with a positive sign instead. The precise slope depends on the model of planet formation and the composition of the planets. In Figure~\ref{fig:theory}, we compare the observed slope with that of late formation in a gas-poor environemnt \citep{lopez2016}, and with different models of photo-evaporation \cite{owen2017}. Because the models use the maximum radius at the bottom of the valley, we compare them to the lower parallel support vector of Figure~\ref{fig:svc}. We find that our slope is consistent at $2\sigma$ with the more complex models, including recombination and X-ray evaporation, and inconsistent with the steeper slope predicted for pure energy-limited evaporation \citep{owen2017}. Finally, it is clear from Figure~\ref{fig:theory} that the location of the photo-evaporation valley is more consistent with iron-rich cores than with icy cores. This was previously pointed out by \cite{owen2017} and \cite{jin2018}, on the condition that the observed valley is indeed primarily caused by photo-evaporation -- as the measurement of the valley's slope in this work appears to confirm.

We finally note that the presence of a clear gap in radius is evidence of largely homogeneous cores, with compositions similar to that of Earth, as a wide range of different compositions would smear out the radius gap \citep{owen2017}. Indeed, if sub-Neptune planets formed beyond the snowline, they would have large amounts of water and volatile ices \citep{rogers2011}, which may completely eliminate the presence of any radius gap \citep{lopez2013}. The presence of a clear gap can therefore be taken as evidence that the observed planets formed in-situ or near-situ \citep[e.g.][]{chiang2013}, i.e.\ planets that have not migrated from beyond the snowline. This is also consistent with the observation of a desert of planets larger than $1.5~R_\oplus$ at ultra-short periods \citep{lundkvist2016,lopez2017}. The gap observed here is inconsistent with late time migration and suggests a core mass function peaking around $3~M_\oplus$ \citep{owen2017}.

If photo-evaporation is indeed responsible for the observed super-Earths at short periods, this has implications for measuring the frequency of habitable zone Earth-like planets as well. Because such efforts often include planets slightly larger than Earth, or planets around later stellar types, they may include planets that are rocky only as a result of photo-evaporation, or are not rocky at all. This would result in an overestimate of the occurrence of true Earth analogs \citep{lopez2016}, indicating that great care must be taken when extrapolating findings of small planets at short orbital periods to more temperate Earth-sized planets.

\section{Conclusions}
\label{sec:conclusions}

Using a sample of planet host stars characterized with asteroseismology, we derive accurate stellar and planetary radii to investigate the presence, location and shape of a radius valley of planet occurrence. Within our sample of 117 planets, we detected a clear bimodal distribution, with super-Earth planets with radii of $\approx~1.5~R_\oplus$ and sub-Neptune planets with radii of $\approx~2.5~R_\oplus$, separated by a clear valley around $2~R_\oplus$ where very few planets are observed. 

\begin{itemize}
 \item The location of the valley has a decreasing radius as a function of orbital period (see Figure~\ref{fig:gapfit}). This negative slope is consistent with predictions for photo-evaporation, while it is inconsistent with the exclusive late formation of gas-poor rocky planets, which would result in a slope with the opposite sign. Taking into account photo-evaporation will also be important when inferring the occurrence of Earth-like planets in the habitable zone \citep{lopez2016}.
 \item We estimated the location of the valley as a function of orbital period and found it to be at $\log_{10} R = m \log_{10} P + a$ with $m = {-0.09^{+0.02}_{-0.04}} $ and $a = 0.37^{+0.04}_{-0.02}$. This equation can be used to determine whether a planet with known orbital period and planet radius is near to or inside the radius gap. This may be particularly important for the \textit{TESS} satellite \citep{ricker2014}, which is expected to discover many planets in the relevant period-radius regime.
 \item The presence of a clear valley implies a homogeneous core composition of the planets in our sample. Planets with a wide range of core compositions, or planets which have formed beyond the snow line, would wash out the valley \citep{owen2017}. 
 \item When comparing the location of the valley with theoretical models, we find it to be broadly consistent with cores consisting of a significant fraction of iron, while inconsistent with mostly icy cores \citep{owen2017,jin2018}.
\end{itemize}

Determining the radii of planets and their host star is crucial for determining the location and shape of the radius valley. Here, asteroseismology achieves this precision \citep{silvaaguirre2015,lundkvist2016}. An important caveat for this approach is the limited sample size. Future transit surveys such as TESS and PLATO \citep{rauer2014} will lead to a larger sample with accurate parameters, and may allow to further refine the properties of the radius valley. Such a larger sample may also allow a detailed inference of the underlying occurrence rate of planets, which for now remains limited to larger but less accurately determined samples \citep{fulton2017}. 

Finally, because of the relative faintness of most stars observed by \textit{Kepler}, no homogeneous inference of the mass of the planets in our sample is available. Future samples may allow the determination of planet mass and mean density, which would provide further tests for photo-evaporation models.	

\section*{Acknowledgements}

We thank the anonymous referee for helpful comments and suggestions that have improved this manuscript. We thank Alan Heavens for discussions on support vector machines. Funding for the Stellar Astrophysics Centre is provided by The Danish National Research Foundation (Grant DNRF106). I.S. acknowledges funding from the European Research Council (ERC) under the European Union’s Horizon 2020 research and innovation programme under grant agreement No 694513. The research was supported by the ASTERISK project (ASTERoseismic Investigations with SONG and Kepler) funded by the European Research Council (Grant agreement no.: 267864). M.S.L. is supported by The Independent Research Fund Denmark's Sapere Aude program (Grant agreement no.: DFF--5051-00130). This research was made with the use of NASA's Astrophysics Data System and the NASA Exoplanet Archive, which is operated by the California Institute of Technology, under contract with the National Aeronautics and Space Administration under the Exoplanet Exploration Program.




\bibliographystyle{mnras}
\bibliography{references_radiusgap} 

\clearpage

\renewcommand{\arraystretch}{1.6}
\begin{table*}
\begin{tabular}{lcccccccc}
Planet	&	R$_\mathrm{p}$ [R$_\oplus$] & Period [d] & M$_\star$ [M$_\odot$]	&R$_\star$ [R$_\odot$]\\
\hline
Kepler-10b&	    $1.473 \pm 0.026$	&$0.83749026 (29)$	&$0.920_{-0.020}^{+0.010}$	&$1.0662_{-0.0075}^{+0.0069}$\\               
Kepler-10c&	    $2.323 \pm 0.028$	&$45.294292 (97)$	&$0.920_{-0.020}^{+0.010}$	&$1.0662_{-0.0075}^{+0.0069}$\\               
Kepler-23b& 	$1.694 \pm 0.076$	&$7.106995 (73)$	&$1.078_{-0.077}^{+0.077}$	&$1.548_{-0.048}^{+0.048}$	\\               
Kepler-23c& 	$3.12 \pm 0.10$	   &$10.742434 (39)$	&$1.078_{-0.077}^{+0.077}$	&$1.548_{-0.048}^{+0.048}$	\\               
Kepler-23d& 	$2.235 \pm 0.088$	&$15.27429 (17)$	&$1.078_{-0.077}^{+0.077}$	&$1.548_{-0.048}^{+0.048}$	\\               
Kepler-25b& 	$2.702 \pm 0.037$	&$6.2385369 (33)$	&$1.160_{-0.050}^{+0.040}$	&$1.299_{-0.016}^{+0.015}$		\\           
Kepler-25c& 	$5.154 \pm 0.060$	&$12.7203678 (35)$	&$1.160_{-0.050}^{+0.040}$	&$1.299_{-0.016}^{+0.015}$		\\           
Kepler-37b& 	$0.354 \pm 0.014$	&$13.36805 (38)$	&$0.810_{-0.010}^{+0.020}$	&$0.7725_{-0.0063}^{+0.0051}$\\               
Kepler-37c& 	$0.705 \pm 0.012$	&$21.302071 (92)$	&$0.810_{-0.010}^{+0.020}$	&$0.7725_{-0.0063}^{+0.0051}$\\               
Kepler-37d& 	$1.922 \pm 0.024$	&$39.792232 (54)$	&$0.810_{-0.010}^{+0.020}$	&$0.7725_{-0.0063}^{+0.0051}$\\               
Kepler-65b& 	$1.409 \pm 0.017$	&$2.1549156 (25)$	&$1.199_{-0.030}^{+0.030}$	&$1.401_{-0.014}^{+0.013}$		\\           
Kepler-65c& 	$2.571 \pm 0.033$	&$5.8599408 (23)$	&$1.199_{-0.030}^{+0.030}$	&$1.401_{-0.014}^{+0.013}$		\\           
Kepler-65d& 	$1.506 \pm 0.040$	&$8.131231 (21)$	&$1.199_{-0.030}^{+0.030}$	&$1.401_{-0.014}^{+0.013}$		\\           
Kepler-68b& 	$2.354 \pm 0.020$	&$5.3987533 (13)$	&$1.070_{-0.020}^{+0.010}$	&$1.2379_{-0.0051}^{+0.0067}$\\               
Kepler-68c& 	$0.927 \pm 0.025$	&$9.604979 (45)$	&$1.070_{-0.020}^{+0.010}$	&$1.2379_{-0.0051}^{+0.0067}$\\
Kepler-92b& 	$3.65 \pm 0.13$	   &$13.748933 (75)$	&$1.209_{-0.030}^{+0.020}$	&$1.719_{-0.013}^{+0.011}$		\\
Kepler-92c& 	$2.455 \pm 0.053$	&$26.72311 (19)$	&$1.209_{-0.030}^{+0.020}$	&$1.719_{-0.013}^{+0.011}$		\\
Kepler-92d& 	$2.067 \pm 0.056$	&$49.3568 (24)$		&$1.209_{-0.030}^{+0.020}$	&$1.719_{-0.013}^{+0.011}$		\\
Kepler-100b&	$1.305 \pm 0.030$	&$6.887037 (47)$	&$1.109_{-0.020}^{+0.020}$	&$1.5131_{-0.011}^{+0.0093}$		\\
Kepler-100c&	$2.221 \pm 0.022$	&$12.815909 (26)$	&$1.109_{-0.020}^{+0.020}$	&$1.5131_{-0.011}^{+0.0093}$		\\
Kepler-100d&	$1.514 \pm 0.034$	&$35.33313 (43)$	&$1.109_{-0.020}^{+0.020}$	&$1.5131_{-0.011}^{+0.0093}$	\\
Kepler-103b&	$3.476 \pm 0.039$	&$15.965316 (18)$	&$1.099_{-0.030}^{+0.019}$	&$1.455_{-0.013}^{+0.024}$		\\
Kepler-103c&	$5.319 \pm 0.052$	&$179.6133 (47)$	&$1.099_{-0.030}^{+0.019}$	&$1.450_{-0.009}^{+0.009}$		\\
Kepler-107b&	$1.581 \pm 0.056$	&$3.180026 (12)$	&$1.142_{-0.068}^{+0.068}$	&$1.411_{-0.047}^{+0.047}$\\
Kepler-107c&	$1.664 \pm 0.065$	&$4.901441 (30)$	&$1.142_{-0.068}^{+0.068}$	&$1.411_{-0.047}^{+0.047}$\\
Kepler-107d&	$1.064 \pm 0.062$	&$7.95825 (11)$		&$1.142_{-0.068}^{+0.068}$	&$1.411_{-0.047}^{+0.047}$	\\
Kepler-107e&	$2.92 \pm 0.10$	   &$14.749176 (34)$	&$1.142_{-0.068}^{+0.068}$	&$1.411_{-0.047}^{+0.047}$	\\
Kepler-108b&	$9.56 \pm 0.53$	   &$49.18354 (18)$	    &$1.377_{-0.089}^{+0.089}$	&$2.19_{-0.12}^{+0.12}$		\\
Kepler-108c&	$8.23 \pm 0.47$    	&$190.3214$ (n/a)	&$1.377_{-0.089}^{+0.089}$	&$2.19_{-0.12}^{+0.12}$		\\
Kepler-109b&	$2.338 \pm 0.034$	&$6.4816370 (80)$	&$1.069_{-0.040}^{+0.040}$	&$1.339_{-0.015}^{+0.017}$		\\
Kepler-109c&	$2.634 \pm 0.043$	&$21.222620 (30)$	&$1.069_{-0.040}^{+0.040}$	&$1.339_{-0.015}^{+0.017}$		\\
Kepler-126b&	$1.439 \pm 0.020$	&$10.495634 (30)$	&$1.148_{-0.049}^{+0.051}$	&$1.345_{-0.018}^{+0.015}$	\\
Kepler-126c&	$1.498 \pm 0.062$	&$21.86964 (10)$	&$1.148_{-0.049}^{+0.051}$	&$1.345_{-0.018}^{+0.015}$	\\
Kepler-126d&	$2.513 \pm 0.031$	&$100.28208 (41)$	&$1.148_{-0.049}^{+0.051}$	&$1.345_{-0.018}^{+0.015}$	\\
Kepler-127b&	$1.52 \pm 0.13$    	&$14.43577 (10)$	&$1.240_{-0.086}^{+0.086}$	&$1.359_{-0.035}^{+0.035}$	\\
Kepler-127c&	$2.389 \pm 0.067$	&$29.39344 (17)$	&$1.240_{-0.086}^{+0.086}$	&$1.359_{-0.035}^{+0.035}$	\\
Kepler-127d&	$2.668 \pm 0.084$	&$48.62997 (57)$	&$1.240_{-0.086}^{+0.086}$	&$1.359_{-0.035}^{+0.035}$	\\
\end{tabular}
\caption{Stellar and planetary parameters of the objects in our sample. Parameters are taken from \protect\cite{vaneylen2015} and \protect\cite{vaneylen2017}, while stellar parameters are originally listed in \protect\cite{huber2013},  \protect\cite{silvaaguirre2015}, and \protect\cite{lundkvist2016}. A full version of this table is available online.\label{tab:parameters}}
\end{table*}

\begin{table*}
\begin{tabular}{lcccccccc}
Kepler-129b&	$2.409 \pm 0.040$	&$15.791619 (53)$	&$1.159_{-0.030}^{+0.030}$	&$1.649_{-0.014}^{+0.012}$		\\
Kepler-129c&	$2.522 \pm 0.066$	&$82.1908$ (n/a)	&$1.159_{-0.030}^{+0.030}$	&$1.649_{-0.014}^{+0.012}$		\\
Kepler-130b&	$0.976 \pm 0.045$	&$8.45725 (11)$		&$0.934_{-0.059}^{+0.059}$	&$1.127_{-0.033}^{+0.033}$	\\
Kepler-130c&	$2.811 \pm 0.084$	&$27.508686 (37)$	&$0.934_{-0.059}^{+0.059}$	&$1.127_{-0.033}^{+0.033}$	\\
Kepler-130d&	$1.31 \pm 0.13$    	&$87.5211 (24)$		&$0.934_{-0.059}^{+0.059}$	&$1.127_{-0.033}^{+0.033}$	\\
Kepler-145b&	$2.56 \pm 0.28$    	&$22.95102 (23)$	&$1.419_{-0.030}^{+0.030}$	&$1.887_{-0.014}^{+0.012}$		\\
Kepler-145c&	$3.92 \pm 0.11$    	&$42.88254 (15)$	&$1.419_{-0.030}^{+0.030}$	&$1.887_{-0.014}^{+0.012}$		\\
Kepler-197b&	$1.064 \pm 0.038$	&$5.599293 (39)$	&$0.922_{-0.059}^{+0.059}$	&$1.120_{-0.033}^{+0.033}$	\\
Kepler-197c&	$1.208 \pm 0.048$	&$10.349711 (54)$	&$0.922_{-0.059}^{+0.059}$	&$1.120_{-0.033}^{+0.033}$	\\
Kepler-197d&	$1.244 \pm 0.049$	&$15.67787 (13)$	&$0.922_{-0.059}^{+0.059}$	&$1.120_{-0.033}^{+0.033}$	\\
Kepler-197e&	$0.983 \pm 0.048$	&$25.2097 (14)$		&$0.922_{-0.059}^{+0.059}$	&$1.120_{-0.033}^{+0.033}$	\\
Kepler-278b&	$4.59 \pm 0.26$    	&$30.15856 (91)$	&$1.298_{-0.076}^{+0.076}$	&$2.935_{-0.066}^{+0.066}$	\\
Kepler-278c&	$3.31 \pm 0.12$    	&$51.0851 (35)$		&$1.298_{-0.076}^{+0.076}$	&$2.935_{-0.066}^{+0.066}$	\\
Kepler-338b&	$2.58 \pm 0.13$    	&$13.72699 (47)$	&$1.142_{-0.084}^{+0.084}$	&$1.735_{-0.082}^{+0.082}$	\\
Kepler-338c&	$2.48 \pm 0.14$    	&$24.31168 (87)$	&$1.142_{-0.084}^{+0.084}$	&$1.735_{-0.082}^{+0.082}$	\\
Kepler-338d&	$2.66 \pm 0.15$    	&$44.4287 (16)$		&$1.142_{-0.084}^{+0.084}$	&$1.735_{-0.082}^{+0.082}$	\\
Kepler-338e&	$1.587 \pm 0.083$	&$9.34149 (40)$		&$1.142_{-0.084}^{+0.084}$	&$1.735_{-0.082}^{+0.082}$	\\
Kepler-444b   &	$0.381 \pm 0.021$	&$3.600125 (28)$	&$0.740_{-0.010}^{+0.010}$	&$0.7492_{-0.0040}^{+0.0046}$\\
Kepler-444c&	$0.490 \pm 0.024$	&$4.545817 (44)$	&$0.740_{-0.010}^{+0.010}$	&$0.7492_{-0.0040}^{+0.0046}$\\
Kepler-444d &	$0.530 \pm 0.025$	&$6.189512 (54)$	&$0.740_{-0.010}^{+0.010}$	&$0.7492_{-0.0040}^{+0.0046}$\\
Kepler-444e &	$0.533 \pm 0.019$	&$7.74350 (10)$		&$0.740_{-0.010}^{+0.010}$	&$0.7492_{-0.0040}^{+0.0046}$\\
Kepler-444f &	$0.679 \pm 0.008$	&$9.740529 (36)$	&$0.740_{-0.010}^{+0.010}$	&$0.7492_{-0.0040}^{+0.0046}$\\	   
Kepler-449b&	$2.056 \pm 0.069$	&$12.58242 (27)$	&$0.969_{-0.053}^{+0.053}$	&$1.467_{-0.033}^{+0.033}$	\\
Kepler-449c&	$2.764 \pm 0.086$	&$33.6727 (10)$		&$0.969_{-0.053}^{+0.053}$	&$1.467_{-0.033}^{+0.033}$	\\
Kepler-450b   &	$6.14 \pm 0.33$	    &$28.454851 (25)$	&$1.346_{-0.084}^{+0.084}$	&$1.570_{-0.085}^{+0.085}$	\\
Kepler-450c   &	$2.62 \pm 0.14$ 	&$15.413135 (85)$	&$1.346_{-0.084}^{+0.084}$	&$1.570_{-0.085}^{+0.085}$	\\
Kepler-450d   &	$0.837 \pm 0.068$	&$7.51464 (23)$		&$1.346_{-0.084}^{+0.084}$	&$1.570_{-0.085}^{+0.085}$	\\
KOI-5b      &	$7.87 \pm 0.14$   	&$4.78032767 (84)$	&$1.199_{-0.030}^{+0.020}$	&$1.795_{-0.014}^{+0.015}$	\\
KOI-5c      &	$0.642 \pm 0.061$	&$7.05174 (13)$		&$1.199_{-0.030}^{+0.020}$	&$1.795_{-0.014}^{+0.015}$	\\
TrES-2&	    $13.21 \pm 0.28$	&$ 2.47061340 (2) $    	&$0.97^{+0.08}_{-0.08}$	&$0.96^{+0.02}_{-0.02}$	\\
HAT-P-7&	$16.88 \pm 0.26$	&$ 2.20473543 (3) $    	&$1.55^{+0.10}_{-0.10}$	&$1.99^{+0.03}_{-0.03}$	\\
HAT-P-11&	$4.887 \pm 0.065$	&$ 4.88780240 (15)$  	&$0.86^{+0.06}_{-0.06}$	&$0.76^{+0.01}_{-0.01}$	\\
Kepler-4&	$4.22 \pm 0.12$   	&$ 3.21367134 (91)$    	&$1.09^{+0.07}_{-0.07}$	&$1.55^{+0.04}_{-0.04}$	\\
Kepler-410&	$2.786 \pm 0.045$	&$17.833613 (47) $    	&$1.22^{+0.07}_{-0.07}$	&$1.35^{+0.02}_{-0.02}$	\\
Kepler-93&	$1.477 \pm 0.033$	&$ 4.72673930 (86)$    	&$0.89^{+0.07}_{-0.07}$	&$0.91^{+0.02}_{-0.02}$	\\
K00075.01&	$10.72 \pm 0.29$   &$105.88162 (75)$     	&$1.32^{+0.07}_{-0.07}$	&$2.58^{+0.07}_{-0.07}$	\\
Kepler-22&	$1.806 \pm 0.029$  &$289.8655 (19)$      	&$0.85^{+0.05}_{-0.05}$	&$0.83^{+0.01}_{-0.01}$	\\
K00092.01&	$3.00 \pm 0.13$	    &$65.70453 (17)$      	&$1.08^{+0.11}_{-0.11}$	&$1.05^{+0.03}_{-0.03}$	\\
Kepler-7&	$17.68 \pm 0.36$	&$ 4.8854862 (12)$   	&$1.28^{+0.07}_{-0.07}$	&$1.97^{+0.04}_{-0.04}$	\\
Kepler-14&	$12.87 \pm 0.26$	&$ 6.7901237 (20)$  	&$1.34^{+0.08}_{-0.08}$	&$2.02^{+0.04}_{-0.04}$	\\
\end{tabular}
\contcaption{Stellar and planetary parameters of the objects in our sample. Parameters are taken from \protect\cite{vaneylen2015} and \protect\cite{vaneylen2017}, while stellar parameters are originally listed in \protect\cite{huber2013},  \protect\cite{silvaaguirre2015}, and \protect\cite{lundkvist2016}. A full version of this table is available online.}
\end{table*}

\begin{table*}
\begin{tabular}{lcccccccc}

Kepler-464&	$3.44 \pm 0.10$ 	&$ 7.257038 (45)$      	&$1.2^{+0.08}_{-0.08}$	&$1.6^{+0.04}_{-0.04}$	&\\
Kepler-467&	$2.26 \pm 0.09$ 	&$24.99337 (21)$       	&$1.01^{+0.07}_{-0.07}$	&$1.36^{+0.04}_{-0.04}$	&\\
Kepler-95&	$3.290 \pm 0.094$	&$11.5230844 (97)$     	&$1.12^{+0.08}_{-0.08}$	&$1.45^{+0.04}_{-0.04}$	&\\
Kepler-506&	$3.088 \pm 0.082$	&$ 6.8834081 (26)$    	&$1.23^{+0.1}_{-0.1}$	&$1.2^{+0.03}_{-0.03}$	&\\
Kepler-96&	$2.647 \pm 0.088$	&$16.2384819 (93)$     	&$1.03^{+0.1}_{-0.1}$	&$0.94^{+0.03}_{-0.03}$	&\\
K00268.01&	$3.043 \pm 0.076$  &$110.38099648 (n/a)$	&$1.2^{+0.07}_{-0.07}$	&$1.36^{+0.03}_{-0.03}$	&\\
K00269.01&	$1.549 \pm 0.047$	&$18.01181 (12)$       	&$1.33^{+0.08}_{-0.08}$	&$1.45^{+0.02}_{-0.02}$	&\\
Kepler-454&	$2.38 \pm 0.094$	&$10.573754 (12)$   	&$1.15^{+0.11}_{-0.11}$	&$1.1^{+0.03}_{-0.03}$	&\\
Kepler-509&	$2.67 \pm 0.20$ 	&$41.746009 (97) $     	&$1.05^{+0.07}_{-0.07}$	&$1.19^{+0.02}_{-0.02}$	&\\
K00280.01&	$2.190 \pm 0.068$	&$11.872877 (11)$      	&$1.03^{+0.09}_{-0.09}$	&$1.04^{+0.02}_{-0.02}$	&\\
Kepler-510&	$2.350 \pm 0.077$	&$19.556464 (62)$     	&$0.81^{+0.11}_{-0.11}$	&$1.38^{+0.04}_{-0.04}$	&\\
K00288.01&	$3.208 \pm 0.055$	&$10.275375 (31)$      	&$1.41^{+0.08}_{-0.08}$	&$2.09^{+0.03}_{-0.03}$	&\\
K00319.01&	$10.36 \pm 0.25$	&$46.15113 (37)$    	&$1.29^{+0.06}_{-0.06}$	&$2.08^{+0.04}_{-0.04}$	&\\
K00367.01&	$4.72 \pm 0.15$  	&$31.578671 (12)$      	&$1.11^{+0.09}_{-0.09}$	&$1.03^{+0.03}_{-0.03}$	&\\
Kepler-540&	$2.947 \pm 0.064$  &$172.70681 (75)$    	&$0.88^{+0.06}_{-0.06}$	&$1.15^{+0.02}_{-0.02}$	&\\
Kepler-643&	$11.29 \pm 0.78$	&$16.338888 (59)$  	    &$1.27^{+0.22}_{-0.22}$	&$2.78^{+0.19}_{-0.19}$	&\\
K00974.01&	$2.601 \pm 0.060$	&$53.50593 (20) $      	&$1.21^{+0.08}_{-0.08}$	&$1.85^{+0.04}_{-0.04}$	&\\
Kepler-21&	$1.707 \pm 0.043$	&$ 2.7858219 (84)$     	&$1.27^{+0.08}_{-0.08}$	&$1.85^{+0.03}_{-0.03}$	&\\
Kepler-805&	$2.611 \pm 0.095$	&$30.8633 (10)$       	&$1.08^{+0.07}_{-0.07}$	&$1.59^{+0.03}_{-0.03}$	&\\
Kepler-432&	$14.7 \pm 2.1$  	&$52.5019 (11)$        	&$1.69^{+0.6}_{-0.6}$	&$4.51^{+0.63}_{-0.63}$	&\\
Kepler-815&	$4.98 \pm 0.60$ 	&$ 8.57522 (22)$       	&$1.69^{+0.5}_{-0.5}$	&$3.88^{+0.43}_{-0.43}$	&\\
Kepler-407&	$1.141 \pm 0.041$	&$ 0.6693127 (20)$	    &$1.02^{+0.07}_{-0.07}$	&$1.02^{+0.02}_{-0.02}$	&\\
Kepler-408&	$0.689 \pm 0.017$	&$ 2.465024 (17)$	    &$1.02^{+0.07}_{-0.07}$	&$1.21^{+0.02}_{-0.02}$	&\\
Kepler-907&	$1.403 \pm 0.081$	&$15.86631 (51)$	    &$0.99^{+0.08}_{-0.08}$	&$1.34^{+0.03}_{-0.03}$	&\\
Kepler-910&	$0.828 \pm 0.049$	&$ 2.364388 (32)$	    &$1.29^{+0.09}_{-0.09}$	&$1.5^{+0.03}_{-0.03}$	&\\
Kepler-911&	$2.44 \pm 0.11$ 	&$20.30895 (73)$       	&$1.22^{+0.08}_{-0.08}$	&$1.93^{+0.04}_{-0.04}$	&\\
Kepler-997&	$1.304 \pm 0.072$	&$ 2.707295 (36)$       &$1.09^{+0.18}_{-0.18}$	&$1.48^{+0.07}_{-0.07}$	&\\
Kepler-1002&$1.609 \pm 0.046$	&$ 4.336422 (31)$     	&$1.18^{+0.07}_{-0.07}$	&$1.54^{+0.02}_{-0.02}$	&\\
Kepler-409&	$1.148 \pm 0.048$	&$68.95825 (29)$      	&$0.95^{+0.08}_{-0.08}$	&$0.9^{+0.02}_{-0.02}$	&\\
K01962.01&	$2.51 \pm 0.14$  	&$32.85861 (58)$       	&$1.04^{+0.07}_{-0.07}$	&$1.5^{+0.04}_{-0.04}$	&\\
K01964.01&	$0.668 \pm 0.029$	&$ 2.2293226 (85)$     	&$0.93^{+0.11}_{-0.11}$	&$0.88^{+0.03}_{-0.03}$	&\\
Kepler-1219&$3.42 \pm 0.37$ 	&$16.1046 (12)$        	&$1.38^{+0.5}_{-0.5}$	&$2.68^{+0.25}_{-0.25}$	&\\
K02462.01&	$1.491 \pm 0.083$	&$12.14533 (70)$        &$1.19^{+0.1}_{-0.1}$	&$1.71^{+0.04}_{-0.04}$	&\\
Kepler-1274&$1.441 \pm 0.071$	&$ 6.98156 (21)$	     &$1.38^{+0.07}_{-0.07}$	&$2.16^{+0.04}_{-0.04}$	&\\
Kepler-1298&$1.588 \pm 0.089$	&$ 7.12836 (47)$	     &$1.37^{+0.17}_{-0.17}$	&$2.16^{+0.07}_{-0.07}$	&\\
K02706.01&	$1.797 \pm 0.082$	&$ 3.097597 (22)$	     &$1.26^{+0.18}_{-0.18}$	&$1.86^{+0.08}_{-0.08}$	&\\
Kepler-1392&$0.684 \pm 0.052$	&$ 2.128229 (24)$	    &$0.99^{+0.15}_{-0.15}$	&$1.3^{+0.06}_{-0.06}$	&\\
K02801.01&	$0.870 \pm 0.061$	&$ 6.99180 (16)$	  &$1.12^{+0.17}_{-0.17}$	&$1.45^{+0.06}_{-0.06}$	&\\
Kepler-1394&$1.04 \pm 0.11$ 	&$ 3.93800 (32)$	  &$1.51^{+0.21}_{-0.21}$	&$1.98^{+0.08}_{-0.08}$	&\\
K03168.01&	$0.988 \pm 0.076$	&$56.382 (45)$         &$1.03^{+0.16}_{-0.16}$	&$1.55^{+0.07}_{-0.07}$	&\\
\end{tabular}
\contcaption{Stellar and planetary parameters of the objects in our sample. Parameters are taken from \protect\cite{vaneylen2015} and \protect\cite{vaneylen2017}, while stellar parameters are originally listed in \protect\cite{huber2013},  \protect\cite{silvaaguirre2015}, and \protect\cite{lundkvist2016}. A full version of this table is available online.}
\end{table*}

\label{lastpage}
\end{document}